\documentclass[preprintnumbers,twocolumn,superscriptaddress,floats,aps,nofootinbib]{revtex4-1}
\pdfoutput=1
\usepackage{graphicx}
\usepackage{color, graphics}
\usepackage{amsmath,amssymb,amsfonts}
\usepackage{verbatim}   
\usepackage{subfigure}  
\usepackage{acronym}
\usepackage{hyperref}
\usepackage{mathrsfs}
\usepackage{multirow}
 \usepackage{cancel}
 \usepackage{url}
  \usepackage{slashed}
 \usepackage{hyperref}

\def\beq{\begin{equation}\begin{aligned}}
\def\eeq{\end{aligned}\end{equation}}

\begin{document}

\title{Fitting the Galactic Center Gamma-Ray Excess with Cascade Annihilations}

\author{Adam Martin}
%\email{adam.martin.322@nd.edu}
\affiliation{Department of Physics, University of Notre Dame, Notre Dame,
IN, 46556, USA}
\author{Jessie Shelton}
%\email{sheltonj@illinois.edu}
\affiliation{Department of Physics, University of Illinois,
1110 West Green Street, Urbana, IL 61801, USA}
\author{James Unwin}
%\email{junwin@nd.edu}
\affiliation{Department of Physics, University of Notre Dame,  Notre Dame,
IN, 46556, USA}

\begin{abstract}

  The apparent excess of gamma rays in an extended region in the
  direction of the galactic center has a spatial distribution and
  amplitude that are suggestive of dark matter annihilations.  If this
  excess is indeed due to dark matter annihilations, it would indicate
  the presence of both dark matter and an additional particle beyond
  the Standard Model that mediates the interactions between the dark
  matter and Standard Model states.  We introduce reference models
  describing dark matter annihilation to pairs of these new mediators,
  which decouples the SM-mediator coupling from the thermal
  annihilation cross section and easily explains the lack of direct
  detection signals.  We determine the parameter regions that give
  good descriptions of the gamma ray excess for several motivated
  choices of mediator couplings to the SM.  We find fermion dark
  matter with mass 7-26 GeV and a dark vector mediator, or scalar dark
  matter in the 10-50 GeV range (Higgs portal mediator) or 10-65 GeV
  range (gluophilic mediator) can provide a comparable or improved
  fit, compared to the case of direct annihilation.  We demonstrate
  that these models can easily satisfy all constraints from collider
  experiments, direct detection, and cosmology.

\end{abstract}

\maketitle

\section{Introduction}

The existence of some form of dark matter (DM) is well-established
from its gravitational interactions.  The coincidence that the
Standard Model (SM) weak interactions roughly coincide with the
coupling strength required for the thermal freeze-out of a weak scale
particle to account for the observed relic abundance has driven a
multipronged search for non-gravitational signals of DM, in direct
detection experiments, in cosmic ray signals, and at colliders.
Over the last few years an excess of gamma rays over astrophysical
predictions has been observed in the Fermi-LAT data in the direction
of the galactic center (GC)
\cite{Goodenough:2009gk,Hooper:2010mq,Hooper:2011ti,Hooper:2012sr,Abazajian:2012pn,
  Hooper:2013rwa,Gordon:2013vta,Huang:2013pda,Abazajian:2014fta,Daylan:2014rsa}.
The extended and spherically symmetric spatial distribution, the
overall rate, and the shape of the spectrum associated with the excess
are strikingly consistent with an origin in annihilating DM.  Of
course, astrophysical sources of the excess are still a possibility.
Millisecond pulsars have been proposed as a source for the excess
\cite{Abazajian:2010zy,Yuan}; however, such a population would have to
be novel in both its spatial and luminosity distributions
\cite{Hooper:2013rwa,Hooper:2013nhl,Daylan:2014rsa,Huang:2013pda}. In
particular, the fact that the excess appears to extend well beyond the
GC, out to at least around 10$^\circ$, is a challenge for the
millisecond pulsar interpretation.  It has also been suggested that
the excess could be due to cosmic ray interactions with gas and dust
\cite{Goodenough:2009gk,Hooper:2011ti,Abazajian:2012pn,Gordon:2013vta},
but similarly it has been argued that these explanations provide a
poor fit to the data \cite{Macias:2013vya,Linden:2012iv}.

One of the most intriguing features of the GC gamma ray excess is that
if it is indeed due to DM, then it would establish the existence not
only of DM but also of some new mediator particle responsible for its
interaction with the SM.  The spectrum of the excess has been fit well
by DM annihilations $X\overline{X}\to b \bar b$, in which case the DM
mass is $m_{\rm DM} \approx$ 30 GeV; annihilations $X \overline{X}\to
\tau\bar \tau$ provide a less good fit to the data and would point to
lighter masses.  Intriguingly, with these hypotheses for the
annihilation, the cross-section needed to achieve the rate seen in the
excess is strikingly close to the thermal value: $\sigma v\sim
10^{-26}~{\rm cm}^{3}/{\rm s}$ \cite{Daylan:2014rsa}.
However, this mass range for DM, $20 \mathrm{\, GeV} \lesssim m_{\rm
  DM}\lesssim 40$ GeV, has been very well studied by direct detection
experiments.  The two SM particles capable of mediating
$X\overline{X}\to f\bar f$ are the Higgs and the $Z$ boson.  With the
coupling to DM set by the deduced cross-section $X\overline{X}\to
f\bar f$, both the Higgs and the $Z$ would give rise to direct
detection signals vastly in excess of observations. In addition, the
failure to observe $h\to X \overline{X}$ at the LHC independently
eliminates the 125 GeV Higgs as a mediator for the DM annihilations.
Thus some beyond-the-SM field is required to mediate DM annihilations
to the SM, whether heavy \cite{Agrawal:2014una,Alves} or light
\cite{Ipek:2014gua,Izaguirre:2014vva,Boehm:2014hva,Berlin:2014tja}.
Other implications and aspects of model building have been considered in
e.g.~\cite{Okada:2013bna,Modak:2013jya,Hagiwara:2013qya,
Hardy:2014dea,Hooper:2012cw,Boucenna:2011hy,Kyae:2013qna,
Ng:2013xha,Marshall:2011mm,Kong:2014haa}.

In the present work we highlight the possibility that the mediator may
easily be lighter than the DM, in which case DM will
typically annihilate to pairs of on-shell mediators rather than
directly to SM particles. In this case, constraints from LHC and
direct detection experiments can be almost entirely obviated
\cite{Pospelov:2007mp}.  Thermal freeze-out determines the coupling of
DM to the mediators, 
but the coupling of the mediators $\phi$ to the SM is a free
parameter. Whilst the coupling of $\phi$ to the SM is constrained by
collider and direct detection searches on one hand, and by the
requirement that they decay before Big Bang nucleosynthesis (BBN) on
the other, a wide range of values are still permitted by all available data.
This allows direct detection signals to be small while still
maintaining a thermal annihilation rate.  We construct three simple
reference models, coupling the mediators to the SM through the
hypercharge portal, through the Higgs portal, and through the 
gluonic operators $G^a_{\mu\nu} G^{a\mu\nu}$ 
and $G^a_{\mu\nu} \widetilde{G}^{a\mu\nu}$, 
and establish the region of cascade annihilation parameter
space that yields a good explanation for the GC excess.

%%%%%%%%%%%%%%%%%%%%%%%%
\begin{figure}[b!]
\includegraphics[height=45mm]{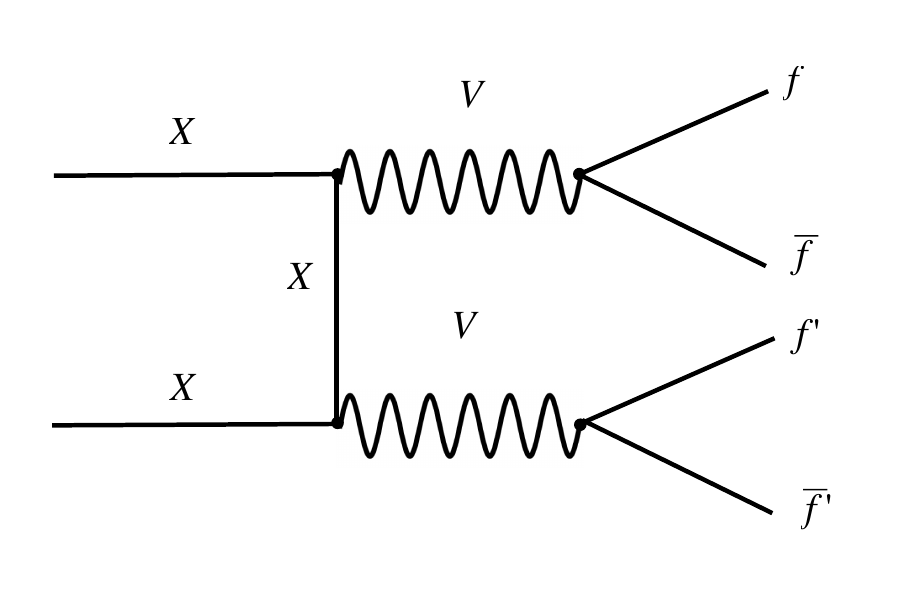}
\caption{Dark matter (particle $X$) cascade annihilation diagram to Standard Model fermion pairs $f\overline{f}$ via intermediate states $V$. \label{cascade}}
\end{figure}
%%%%%%%%%%%%%%%%%%%%%%%%

The paper is structured as follows: In Sect.~\ref{S1} we examine the
photon spectra coming from DM cascade annihilations to SM states and
identify the best-fit parameter regions which reproduce the spectrum
of the inner galaxy gamma ray excess. In Sect.~\ref{S3} - \ref{S5},
we present simple reference models which yield cascade
annihilations capable of describing the excess, and discuss the viable
parameter space. We examine the relevant
limits and discuss the prospect for signals at colliders and direct
detection experiments.  Finally in Sec.~\ref{S6} we conclude.

%%%%%%%%%%%%%%%%%%%%%%%%%%%%%%%%

\begin{figure}[t!]
\includegraphics[height=75mm]{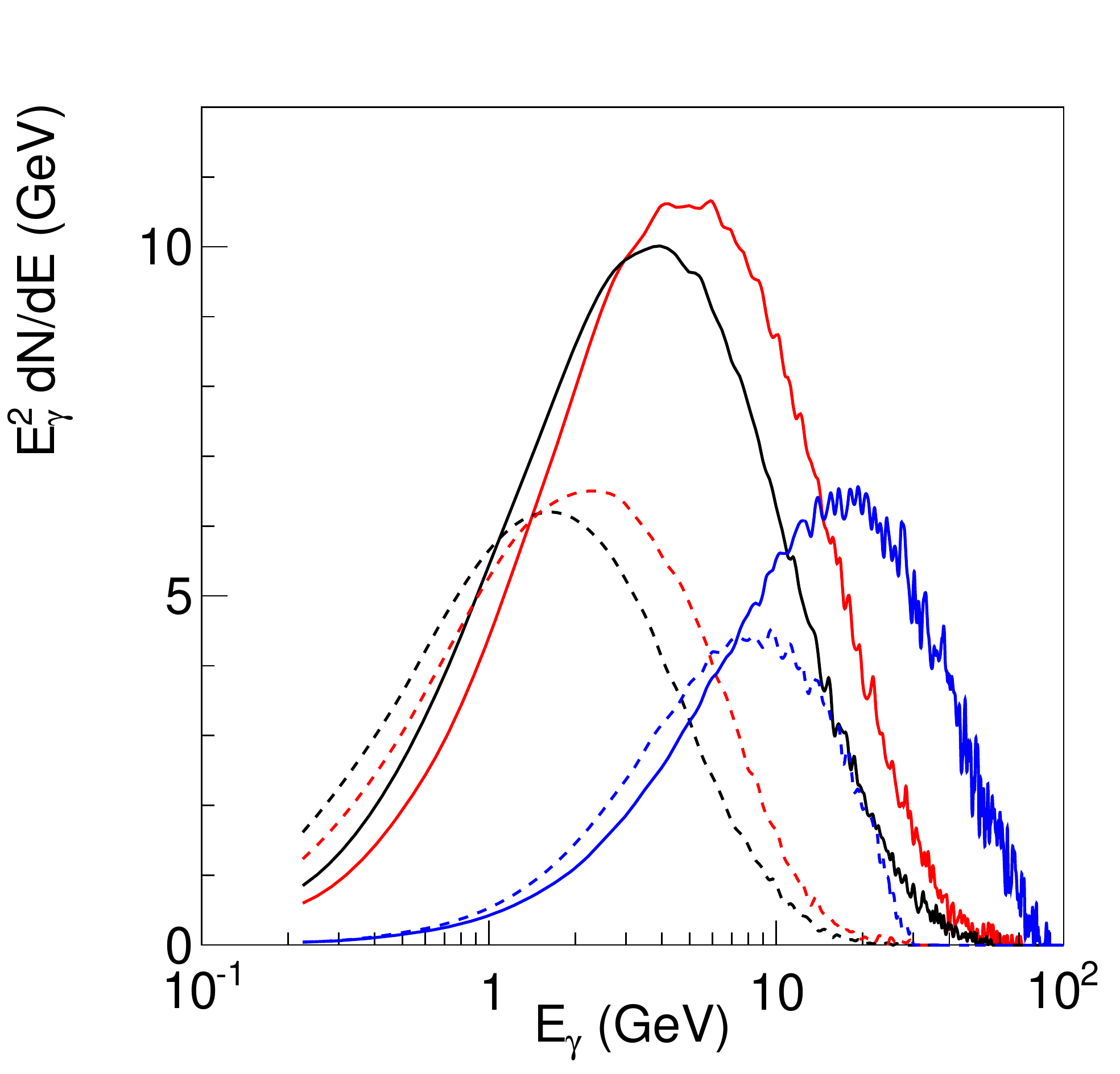}
\caption{
Examples of photon spectra (from {\tt Pythia 8}) assuming DM
  annihilation to $c \overline c$ (red), $b \overline b$ (black),
  $\tau\overline\tau$ (blue).  For direct decays with $m_{\rm DM}=30$ GeV
  (dotted) and cascade decays with $m_{\rm DM}=100$ GeV and $m_V=30$ GeV
  (solid).  
\label{py}}
\end{figure}

%%%%%%%%%%%%%%%%%%%%%%%%%%%%%%%%

%%%%%%%%%%%%%%%%%%%%%%%%%%%%%%%%%%%%%%%%%%%%%%%
\section{Fitting the excess with dark matter cascade annihilations}
\label{S1}
%%%%%%%%%%%%%%%%%%%%%%%%%%%%%%%%%%%%%%%%%%%%%%%

In this section we examine the photon spectra that result from DM
cascade annihilations
\beq
 X \overline{X}\to \phi\phi \to f\bar f f'\bar{f'},
\eeq
as illustrated in Fig.~\ref{cascade}. We study how the spectra vary
with respect to $m_{\rm DM}$ and $m_{\rm med}$, and determine which
combinations of new particle masses and SM final states provide a good
fit to the shape of the spectrum extracted in \cite{Daylan:2014rsa}.

The flux of gamma rays $\Phi(E_\gamma,\psi)$ produced at energy
$E_\gamma$ due to annihilating DM with mass $m_{\rm DM}$, as a function of
angle of observation $\psi$, is given by 
\beq 
 \frac{{\rm d}^2\Phi}{{\rm d}\Omega{\rm d}E_{\gamma}}=\frac{\langle \sigma v
  \rangle}{8\pi \eta m_{\rm DM}^2}\left(\sum_f\frac{{\rm d}N}{{\rm
      d}E_{\gamma}}{\rm Br}_f\right)J(\psi,\gamma)~,
\label{phi}
\eeq
where $\langle \sigma v \rangle$ is the thermally averaged DM
annihilation cross section, ${\rm Br}_f$ is the branching ratio to a
given final state $f$, and the sum runs over all final states. The
flux also depends on whether the DM is complex ($\eta = 2$) or
self-conjugate ($\eta = 1$). Unless stated otherwise, we shall assume
that the DM is non-self-conjugate (e.g.~Dirac) and accordingly take
$\eta=2$.  The factor $J$ is the line-of-sight integral (see
e.g.~\cite{Cirelli:2010xx})
\beq
J(\psi,\gamma)=\int_{\rm LOS} \rho^2(r){\rm d}l~,
\label{LOS}
\eeq
which depends on the distribution of the DM $\rho(r)$ as a function
of radial distance from the GC. Previous studies \cite{Goodenough:2009gk,Hooper:2010mq,Hooper:2011ti,Hooper:2012sr,Abazajian:2012pn,Hooper:2013rwa,Gordon:2013vta,Huang:2013pda,Abazajian:2014fta,Daylan:2014rsa}
have established that the spatial distribution of the gamma ray excess is
well-fit by a generalised Navarro-Frenk-White (NFW) distribution
\cite{Navarro:1995iw,Navarro:1996gj}
\beq
\rho(r)=\rho_0\frac{(r/r_s)^{-\gamma}}{(1+r/r_s)^{3-\gamma}}~,
\label{NFW}
\eeq
with best-fit shape parameter $\gamma \simeq 1.26$
\cite{Daylan:2014rsa}, local DM density $\rho_0=0.3~{\rm GeV}/{\rm
  cm}^3$, and scale radius $r_s$ which we take to be $20$ kpc.  These
values are in agreement with the suggested range in
\cite{Iocco:2011jz}. The normalization of the gamma ray signal depends
on the product $J\langle\sigma v\rangle$. It is important to note that
uncertainties in the astrophysical distribution of DM complicate the
determination of the thermal cross-section.  We will adopt the
absolute normalization of the excess as determined in
\cite{Daylan:2014rsa} together with their best-fit $\rho(r)$, and
report on the deduced cross-section $\langle \sigma v (X
\overline{X}\to \phi \phi) \rangle$ needed to fit the rate.  We
compare this to predictions from specific thermal reference models in
Sects.~\ref{S3} - \ref{S5}.

Out first step is to establish which combinations of masses and final
states yield a good description of the shape of the photon spectrum.
For a given DM mass and final state we determine the photon spectra
using {\tt Pythia 8} \cite{Sjostrand:2007gs}. Example spectra for DM
annihilations to either $b, c$ or $\tau$ final states are plotted
below in Fig.~\ref{py}. We show examples of both direct annihilations
$X\overline{X}\rightarrow f\overline{f}$ and cascade annihilations
$X\overline{X}\rightarrow f\overline{f}f\overline{f}$. Shifting from
direct to cascade annihilations affects both the height and peak
location in the $E^2_{\gamma}\,{\rm d}N/{\rm d}E$ distributions.

%%%%%%%%%%%%%%%%%%%%%%%%%%%%%%%%%%%%%%%%%%%%%%%%%
\begin{figure*}[t!]
\begin{center}
\includegraphics[height=63mm]{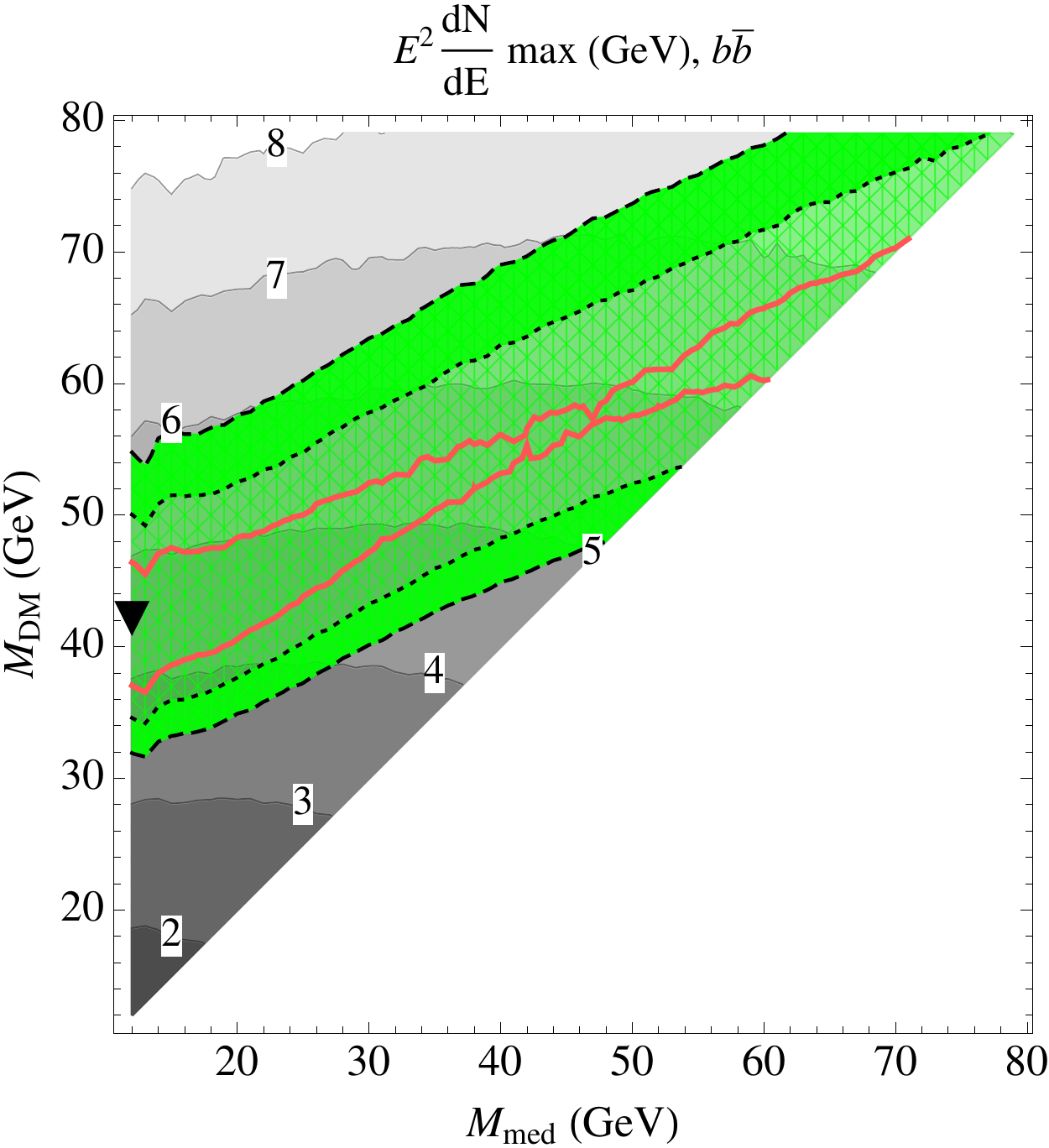}
\includegraphics[height=63mm]{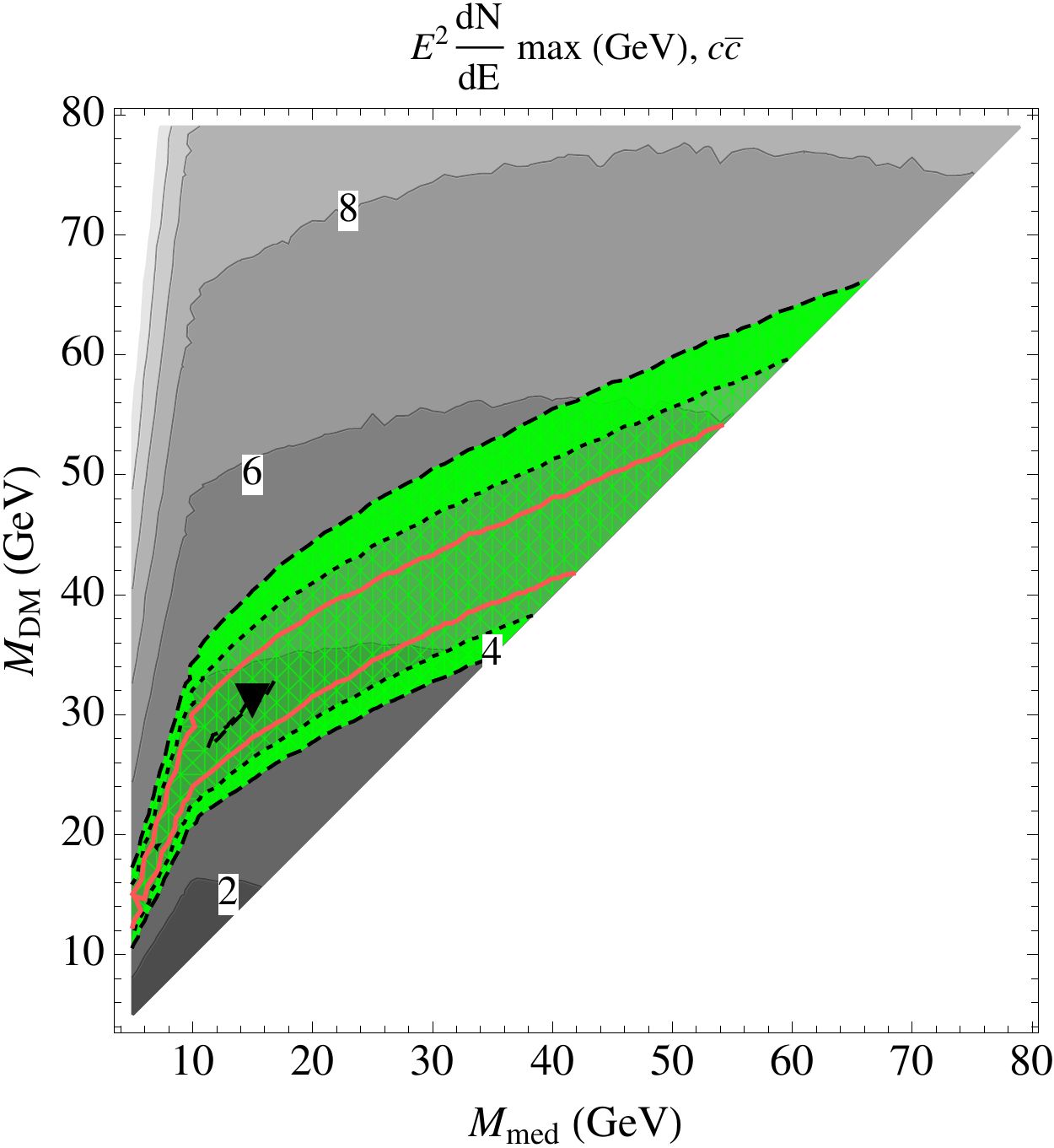}
\includegraphics[height=63mm]{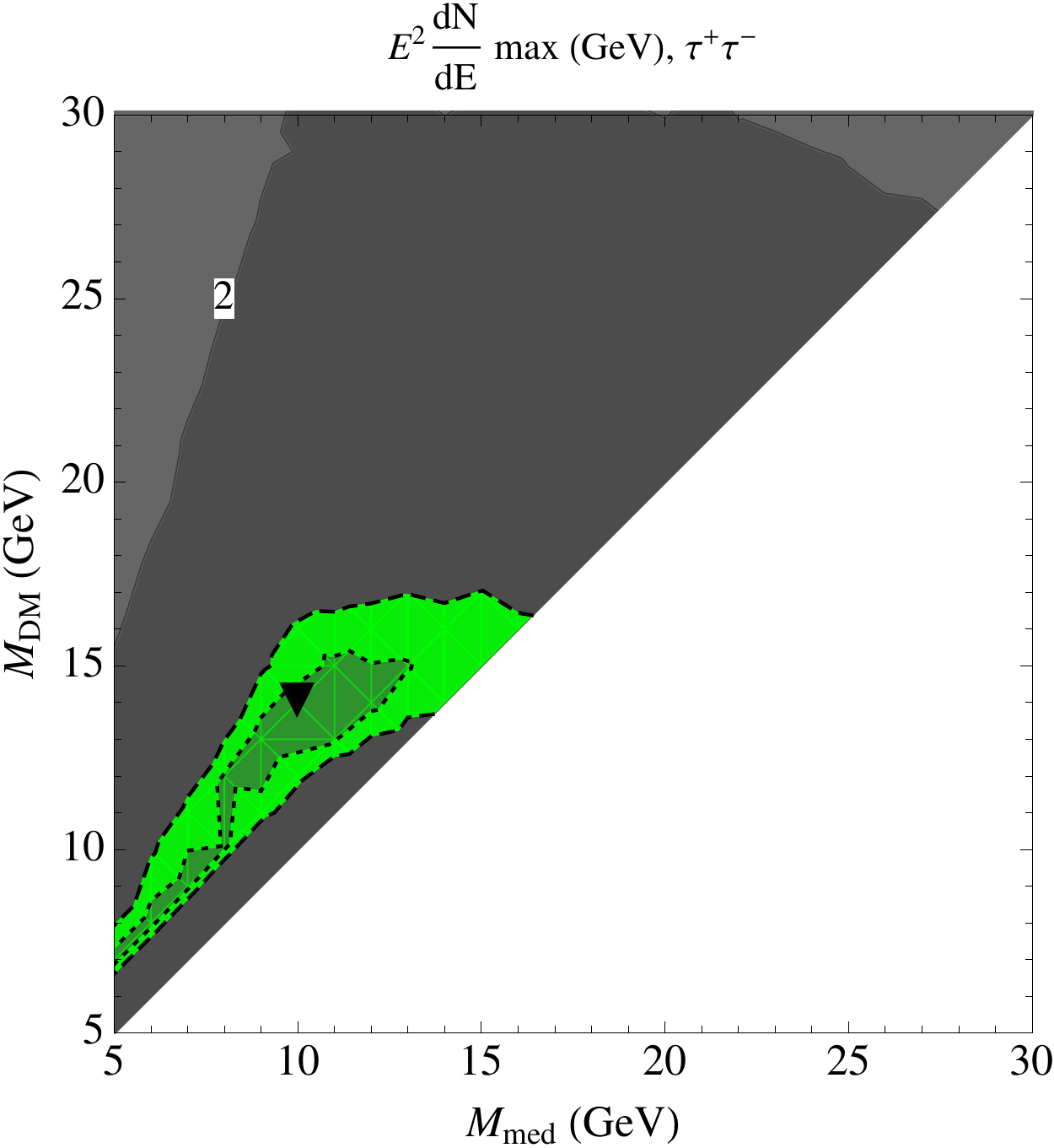}
\caption{ Contour plots of max$\left[E_{\gamma}^2\frac{{\rm d}N}{{\rm
        d}E_{\gamma}}\right]$ as a function of $m_{\rm DM}$ and $m_{\rm med}$
  for cascade decays to $b\overline{b}$, $c\overline{c}$,
  $\tau\overline{\tau}$ (grey contours). For a spectrum which fits the shape of the flux, the corresponding value of
${\rm max} \left[E_\gamma^2~{\rm d}N/{\rm d}E_\gamma\right]$ may
be used to obtain the normalization factor $J\langle \sigma v
\rangle$ required to match the signal.
  Parameter regions that provide a good fit to
  the observed spectrum are obtained are highlighted in green, showing
  contours of $\chi^2$. Parameter space within the red curves
  indicates a better fit than achieved for the case of direct
  annihilations, as identified in \cite{Daylan:2014rsa}. 
The black markers indicate the best fit points for each case.
  Further details are given in the text.
 \label{figB}}
\end{center}
\end{figure*}
%%%%%%%%%%%%%%%%%%%%%%%%%%%%%%%%%%%%%%%%%%%%%%%%%%%%%%%%%%%%%%%%%%%%5

For a given choice of $m_{\rm DM}$, $m_{\rm med}$ and final state, we
fit the resulting photon spectrum to the data shown in
\cite{Daylan:2014rsa}.  To perform this fit, we fix the normalization
of each $E^2_{\gamma}\,{\rm d}N/{\rm d}E$ curve so that the integrated
flux of the cascade model is normalised to the flux reported in
\cite{Daylan:2014rsa}.  We then define a $\chi$-square statistic that
compares the predicted flux from a given cascade model in each bin
${N^{\rm mod}_i}$ to the observed flux $N^{\rm data}_i$, as given in
\cite{Daylan:2014rsa}, and sums over the bins
\beq
\chi^2 = \sum_i\frac{({N^{\rm data}_i} - {N^{\rm mod}_i})^2}{\sigma_i^2}~,
\eeq
where $\sigma_i$ is the error on the data. Contours of the best fit
regions in the $m_{\rm DM}$, $m_{\rm med}$ plane that give good fits
to the shape of the photon spectrum for cascade annihilations to
different final states are shown in Fig.~\ref{figB}. The green
contours show the best fit region according to the $\chi^2$ test\footnote{We give the ratio of $\chi^2$ relative to the ($25-1$) degrees of freedom, corresponding to the 25 data points used in the analysis of \cite{Daylan:2014rsa}.}; the
inner contour corresponds to $\chi^2/\mathrm{d.o.f} =2$, and the
outer contour is $\chi^2/\mathrm{d.o.f} = 3$. The $\chi^2$ contours
are overlaid over grey contours of ${\rm max} \left[E_\gamma^2~{\rm
    d}N/{\rm d}E_\gamma\right]$, normalized to half the number of
photons per annihilation. For spectra that fit the data well, the
${\rm max} \left[E_\gamma^2~{\rm d}N/{\rm d}E_\gamma\right]$ can then
be used to determine the normalization factor $J\langle \sigma v
\rangle$ necessary to fit the observed flux.

The red line in Fig.~\ref{figB} corresponds to a fit of quality
comparable to that found by \cite{Daylan:2014rsa} for direct
annihilations; the best fit DM mass for $X\overline{X}\to
b\overline{b}$ was identified as $m_{\rm DM}\simeq$~32.25 GeV with
$\chi^2/{\rm d.o.f}=1.4.$\footnote{To be consistent, this is the
  $\chi^2$ calculated by our procedure for $ 32.25\,\text{GeV}$ DM
  annihilating to $\bar b b$, not the value quoted
  by~\cite{Daylan:2014rsa}.}
Thus the parameter space which lies within the red curves provides an
improved fit, which is not unexpected as the cascade scenario
introduces additional parameters. The lack of a red curve in the
rightmost panel of Fig.~\ref{figB} means all $(m_{\rm med}, m_{\rm
  DM})$ points lead to spectra with $\chi^2/\mathrm{d.o.f} > 1.4$.

We consider only prompt photons, meaning photons from decay of or
radiation off the final state. Secondary photons arising from the
interaction of DM annihilation products with dust, gas, and magnetic
fields (and, to a lesser extent, with starlight and the CMB) can be
important for accurately extracting and understanding the spectrum of
the DM annihilations \cite{Cirelli:2013mqa}.  The magnitude and
morphology of these secondary contributions depend on the distribution
of matter and radiation in the galaxy. Away from the galactic disk, we
expect the contribution of secondary photons to be relatively
unimportant, and, following \cite{Daylan:2014rsa} we fit to the flux
as determined at $\psi =5^{\circ}$ where bremsstrahlung and other
secondary processes can largely be neglected.

%%%%%%%%%%%%%%%%%%%%%%%%%%%%%%%%%%%%%%%%%%%%%%%%%%%%%%%%%%%%%%%%%%%%%%
\begin{figure*}[t!]
\begin{center}
\includegraphics[height=63mm]{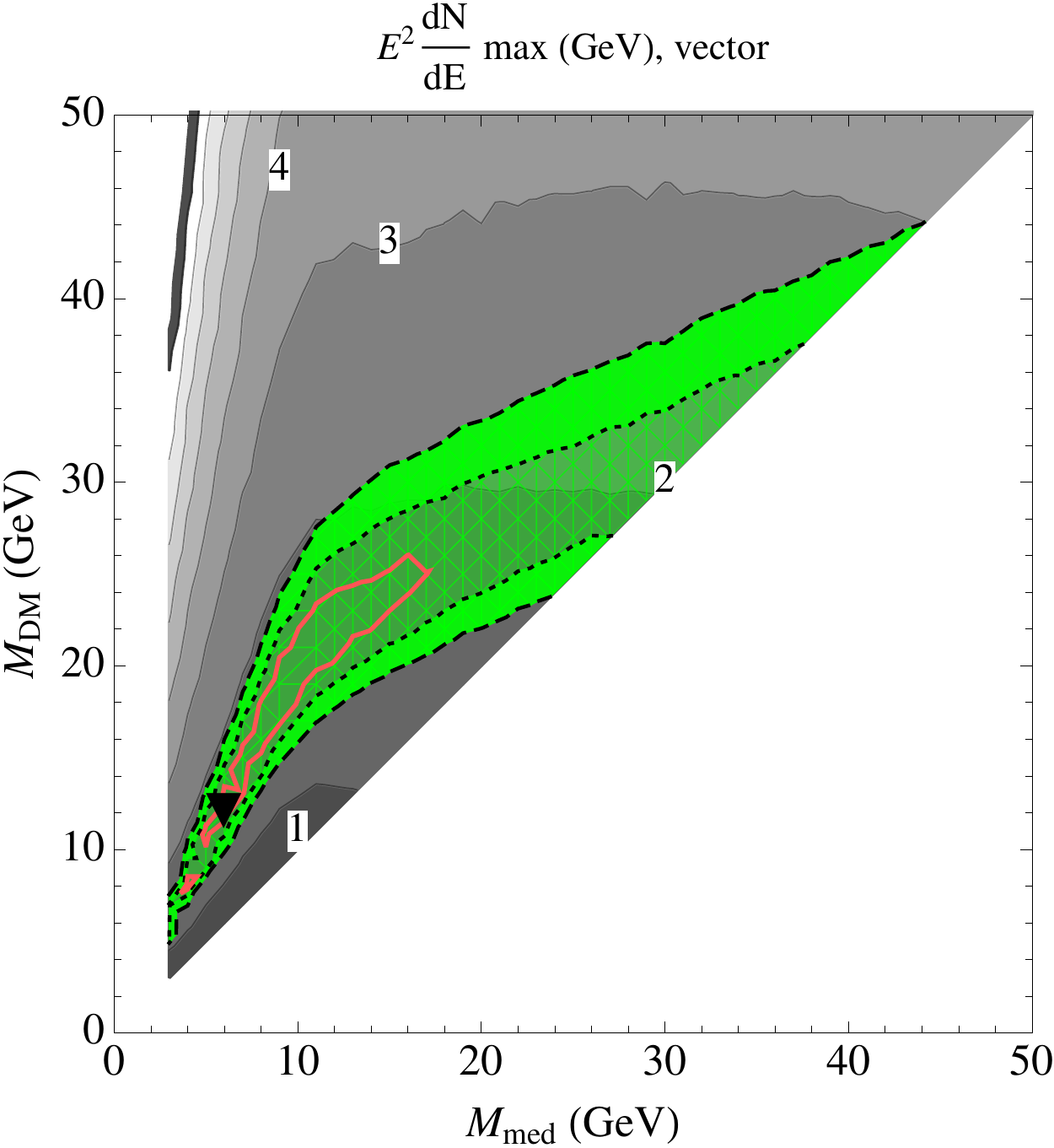}
\includegraphics[height=63mm]{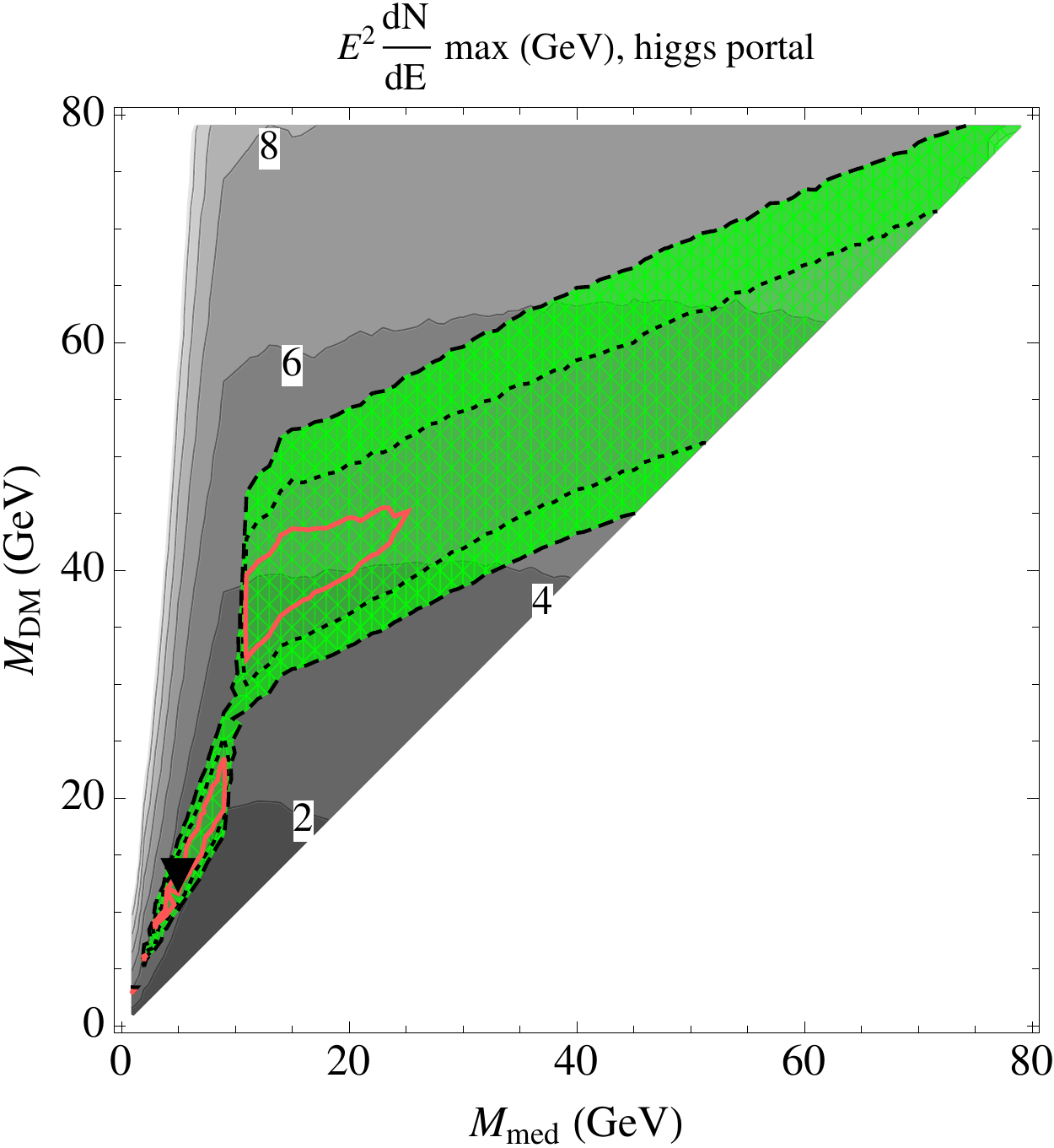}
\includegraphics[height=63mm]{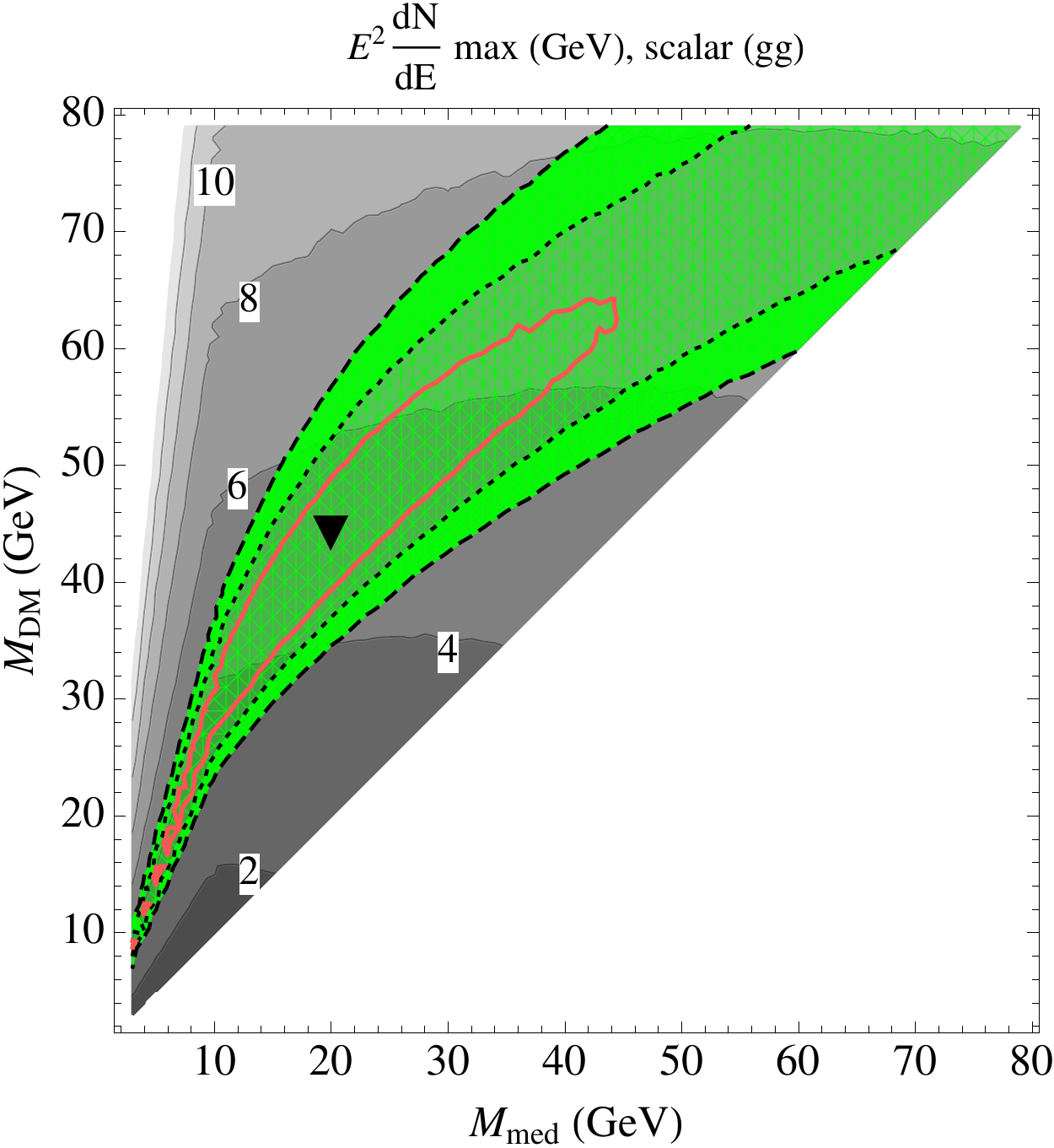}
\caption{As in Fig.~\ref{figB}, for specific choices of mediator. The left panel corresponds to a dark vector mediator $X\overline{X}\rightarrow V_\mu V^\mu$, with $V^\mu$ coupling to the SM through the
hypercharge portal, as discussed in Sect.~\ref{S3}. Cascade annihilation to scalar states $X\overline{X}\rightarrow ss$ which subsequently decay to SM fermions through the Higgs portal, Sect.~\ref{S4}, is shown in the center panel.  The right panel shows annihilation $X\overline{X} \to a a$, with $a\to gg$, as discussed in Sect.~\ref{S5}.   
The black markers indicate the best fit points.
\label{figE}}
\end{center}
\end{figure*}
%%%%%%%%%%%%%%%%%%%%%%%%%%%%%%%%%%%%%%%%%%%%%%%%%%%%%%%%%%%%%%%%%%%%%

\begin{figure}[b!]
\begin{center}
\includegraphics[height=85mm]{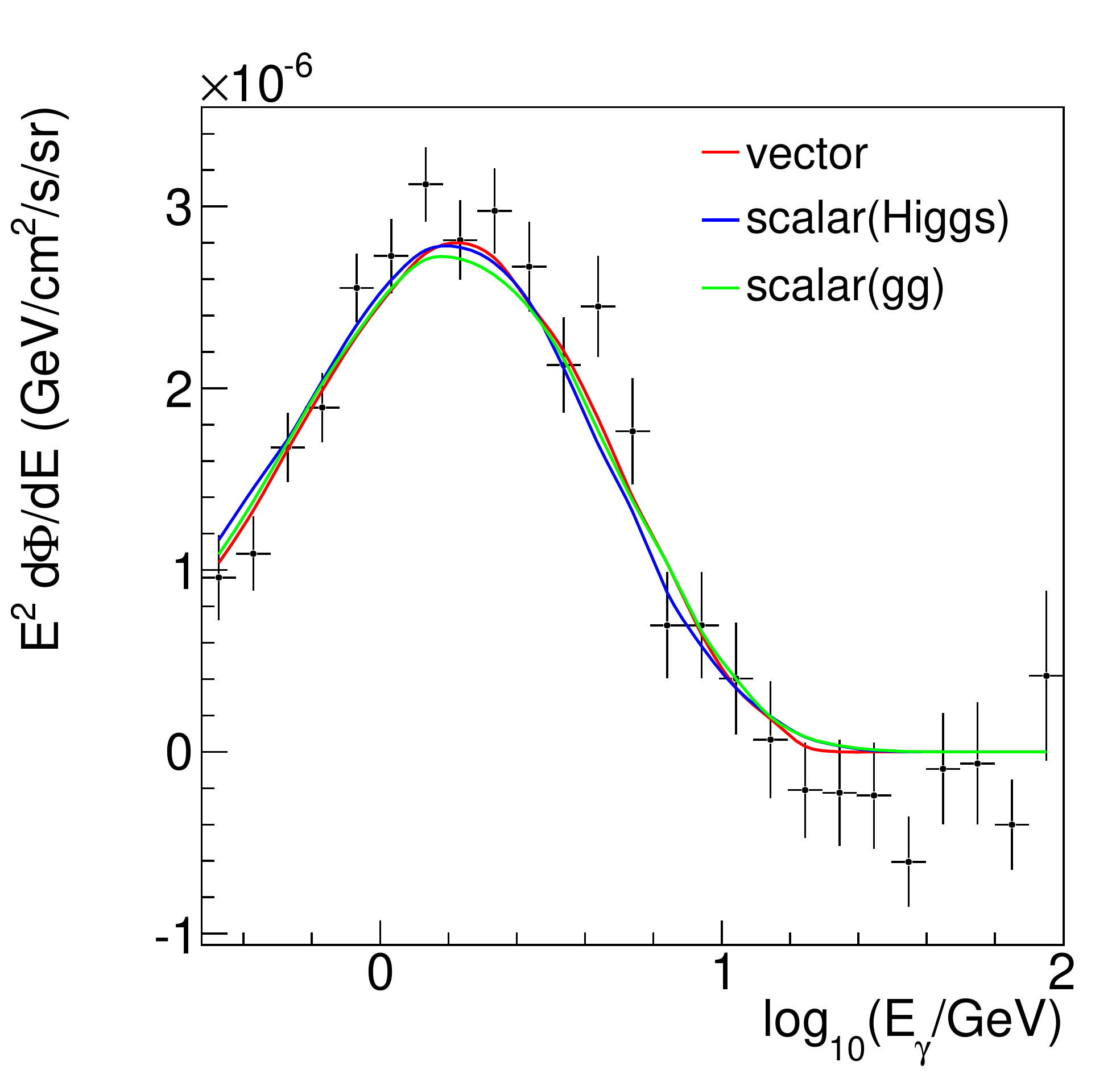}
\caption{
Gamma-ray flux observed in the inner galaxy, as reported by \cite{Daylan:2014rsa}, compared to the best-fit spectra for each model of DM cascade annihilation.
 Red: Dark vector model with $m_{\rm DM}=22$ GeV and $m_{\rm med}=12$ GeV, for DM annihilation cross section of $\langle \sigma v\rangle =3.4\times10^{-26}$ cm${}^3/$s. 
Blue: Scalar decaying via the Higgs portal for $m_{\rm DM}=40$ GeV and $m_{\rm med}=20$ GeV, for  $\langle \sigma v\rangle =4.4\times10^{-26}$ cm${}^3/$s. 
Green: Scalar decaying to gluons with $m_{\rm DM}=60$ GeV and $m_{\rm med}=40$ GeV, for $\langle \sigma v\rangle =6.1\times10^{-26}$ cm${}^3/$s.  
In each case the flux is calculated for angular direction $5^{\circ}$ from GC, for a generalised NFW profile with $\gamma=1.26$ and $\rho_0=0.3~{\rm GeV}/{\rm  cm}^3$.
\label{figF}}
\end{center}
\end{figure}

In Fig.~\ref{figE} we show the corresponding plots for cascade
annihilations to mediators with specific choices of branching
fractions to SM fermions.  We consider three scenarios for the
coupling of the mediator to the SM.  First, we consider the case where
the mediator is a vector boson $V$, coupling to the SM via the
hypercharge portal, $\epsilon\, F_{\mu\nu} B^{\mu\nu}$
\cite{Holdom:1985ag, Galison:1983pa}.  For $m_V \lesssim 20$ GeV, the
branching fractions of the $V$ to SM fermions are approximately
proportional to electric charge; for heavier $m_V$, the branching
fractions begin to become more $Z$-like.  We use the mass-dependent
branching fractions, determined at tree level (see, e.g.,
\cite{Gopalakrishna:2008dv}).  Second, we consider the case where the
mediator is a scalar, which couples to the SM via the Higgs portal,
and has branching fractions given at tree-level by the SM Yukawa
couplings. In our numerical work we use branching fractions computed
by \texttt{HDECAY} \cite{Djouadi:1997yw,Butterworth:2010ym}, including
higher-order corrections and the loop-induced mode $h\to g g$.
Finally, we consider the case where the mediator is a (pseudo-)scalar
coupling to the SM through the higher dimensional operators $a\,
G^a_{\mu\nu} \widetilde{G}^{a\mu\nu}$ or $s\, G^a_{\mu\nu}
G^{a\mu\nu}$, resulting in $a(s)\to g g$. Further, in Fig.~\ref{figF}
we show the corresponding flux for the best fit of each of the cases
studied in Fig.~\ref{figE}, using the preferred astrophysical
parameters identified in \cite{Daylan:2014rsa}.

One expected consequence of fitting the spectrum with cascade
annihilations is the expanded range of $m_{\rm DM}$ that can yield a
good description of the shape. Generically cascade annihilations point
to heavier DM masses than do direct annihilations. The diagonal line
where $m_{\rm DM} = m_{\rm med}$ recovers the spectra for direct
annihilations, $X\overline{X}\to f\overline{f}$, at the mass
\beq
\left. m_{\rm DM}\right|_{\rm direct}
= 0.5 \left. m_{\rm DM}\right|_{\rm cascade}~,
\eeq
with {\em twice} as many photons per event.  This limit is not of
particular interest for describing the observed GC photon excess, as
the thermal cross-section is proportional to $\beta_f =
\sqrt{1-m^2_{\rm med}/m^2_{\rm DM}}$, and becomes velocity-suppressed
in the degenerate limit.  Cascade annihilations are a more compelling
scenario away from the diagonal line.  It is therefore interesting
that in all cases our best-fit regions lie at $m_{\rm med}/m_{\rm
  DM}<1$.
  
Given that the ratio of masses should be $m_{\rm med}/m_{\rm
  DM}\gtrsim0.1$ in order to reproduce the shape of the 1-3 GeV photon
excess, the dark mediator is not in a regime in which Sommerfeld
effects are typically important, see
e.g.~\cite{ArkaniHamed:2008qn}. The DM elastic self-interactions
resulting from mediator exchange are thus not in the range that could
account for the astrophysical small scale structure anomalies, such as
the core-vs-cusp problem, for which self-interacting DM models
substantially lighter mediators have been proposed as a solution
\cite{Spergel:1999mh,Buckley:2009in}.  Conversely, the cascade
annihilation scenarios we consider here are not constrained by limits
on DM-DM self-scattering, such as those arising from the observation
of elliptical halos \cite{Feng:2009hw}.

Another noticeable feature of the left and middle panels of
Fig.~\ref{figE} is the change in shape of the $\chi^2$ contour around
$m_{\rm med} = 10\,\text{GeV}$, where the $\to \bar bb$ final state
becomes accessible.

%%%%%%%%%%%%%%%%%%%%%%%%%%%%%%%%

%%%%%%%%%%%%%%%%%%%%%%%%%%%%%%%%

%%%%%%%%%%%%%%%%%%%%%%%%%%%%%%%%%%%%%%%%%%%%%%%%%%%%%%%%%
\section{Cascade annihilation to vector mediators}
\label{S3}
%%%%%%%%%%%%%%%%%%%%%%%%%%%%%%%%%%%%%%%%%%%%%%%%%%%%%%%%%%%

%%%%%%%%%%%%%%%%%%%%%%%%%%%%%%%%%%%%
\begin{figure*}[t!]
\begin{center}
\includegraphics[height=65mm]{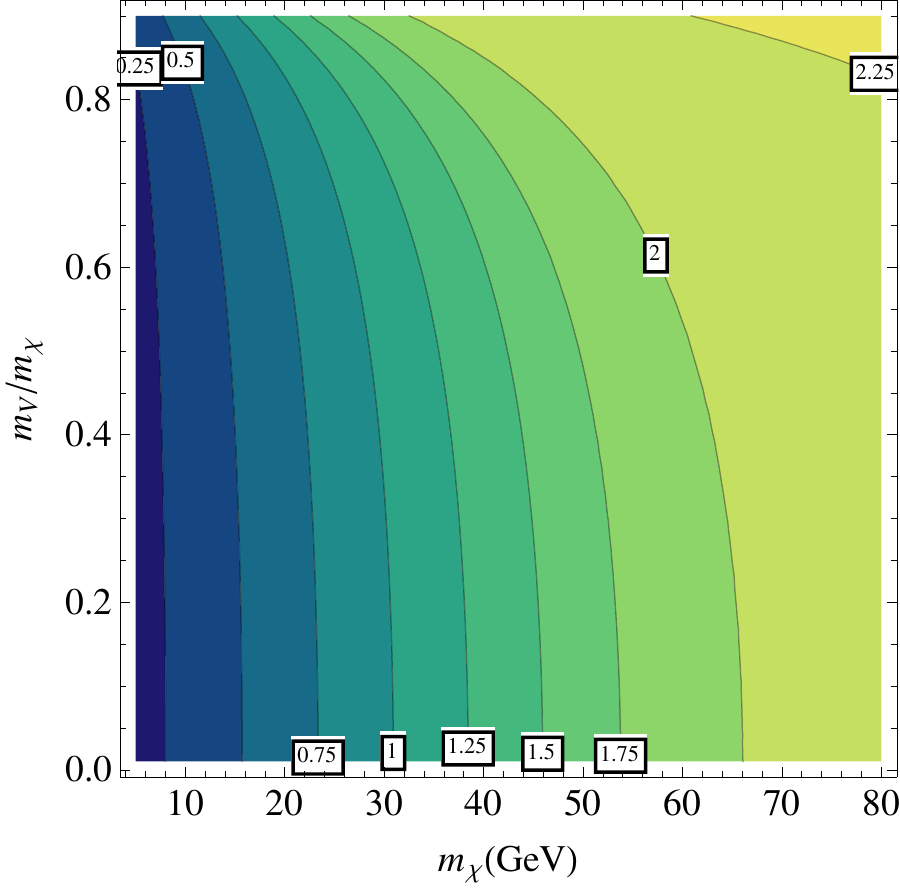}
\hspace{20mm}
\includegraphics[height=65mm]{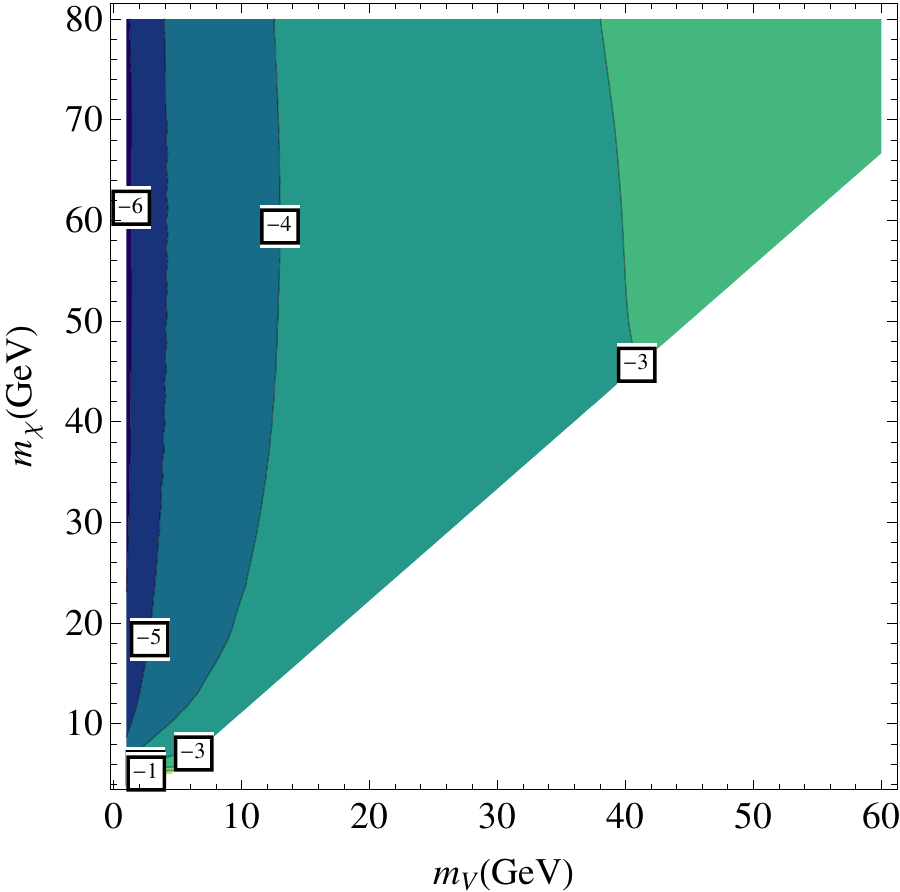}
\caption{Left: The dark gauge coupling required to obtain $\Omega h^2
  = 0.11$ for Dirac fermion DM $\chi$ annihilating to pairs
  of dark gauge bosons $V$, as a function of $m_{\chi}$ and $m_V = x
  m_\chi$ (using the approximated $s$-wave solution to the Boltzmann
  equation). Contours show the required value of $\alpha_D$ in units
  of $10^{-3}$.  Right: The maximum value of the kinetic mixing
  parameter $\epsilon$ allowed by LUX \cite{Akerib:2013tjd}, when the
  dark coupling constant $\alpha_D$ has been set to its thermal relic
  value.  Contours of $\log_{10} \epsilon$ are shown.}
\label{fig:alphaDthermal}
\end{center}
\end{figure*}

%%%%%%%%%%%%%%%%%%%%%%%%%%%%%%%%

In this section we construct an explicit reference model for DM
annihilating to pairs of vector bosons, and discuss direct detection
and LHC constraints.  The regions of DM mass and vector mass that
yield good fits to the GC excess are shown in Fig.~\ref{figE}.

We take the DM to be a Dirac fermion $\chi$, interacting with a
massive dark vector $V_\mu$ with coupling strength $g_D$, and couple
the dark vector to the SM through the hypercharge portal, $\mathcal{L}
= \epsilon\, B_{\mu\nu} V^{\mu\nu}$.  This kinetic mixing couples
$V_\mu$ to SM fermions, thereby destabilizing $V_\mu$, and yielding a
direct detection signal.  The relic abundance of $\chi$ is set by its
annihilation to pairs of vectors, $\chi\bar\chi \to V V$.  This
process has an $s$-wave contribution, and thus yields present-day
annihilation cross-sections of the correct order of magnitude to
explain the GC excess.  Both the direct detection cross-section and
the $V_\mu$ total width are proportional to $\epsilon ^ 2$.  On one
hand, $V_\mu$ must decay sufficiently promptly to avoid disrupting the
predictions of BBN. On the other hand, the direct detection
cross-section must be small enough not to conflict with LUX bounds
\cite{Akerib:2013tjd}.  Thus, the admissible range of mediator-SM
couplings is bounded \cite{Kaplinghat:2013yxa}.

In what follows, we fix the dark coupling constant $g_D$ to the value
that yields the correct thermal relic abundance for $\chi$.  The
thermal annihilation cross-section is
\beq
\langle \sigma v \rangle = \frac{\pi \alpha_D^2 (1-m_V^2/m_\chi^2)^{3/2}}{m_\chi^2(1-m_V^2/2m_\chi^2)^2}
          + \mathcal{O}(v^2).
\eeq
Requiring $\Omega h^2 = 0.112 $ fixes $\alpha_D\equiv g_D^2/4\pi$ for
given $m_V, m_\chi$.  Note we are working in the limit where the Born
approximation holds, i.e.~where $\alpha_D m_\chi/m_V < 1$, and
enhancements to cross-sections at low velocities are negligible.  The
resulting value for $\alpha_D$ is shown in
Fig.~\ref{fig:alphaDthermal} (left), where we have used the standard
analytic approximate solution to the Boltzmann equation assuming
$s$-wave freeze-out.  This is a good approximation for $x\equiv
m_V/m_\chi \lesssim 0.7$ and a not unreasonable approximation for
$0.7\lesssim x\lesssim 0.9$.  In the degenerate limit $m_V\to m_\chi$,
the $s$-wave term in the $v^2$ expansion vanishes, and the resulting
thermal annihilation cross-section is highly velocity-suppressed; we
consider this region of parameter space of less interest for
explaining the excess in the GC.

For the best-fit region of $m_V$, $m_\chi$ indicated in
Fig.~\ref{figE}, exactly solving the Boltzmann equation predicts that
the thermal cross-section today lies in the range
\beq
\label{eq:sigmavtodayV}
\left. \langle \sigma v \rangle\right|_{\mathrm{today}} = (4.3-4.5)\times 10^{-26} \mathrm{cm^3}/\mathrm{s}.
\eeq
%

%%%%%%%%%%%%%%%%%%%%%%%%%%%%%%%%%%%%
\begin{figure*}[t!]
\begin{center}
\includegraphics[height=65mm]{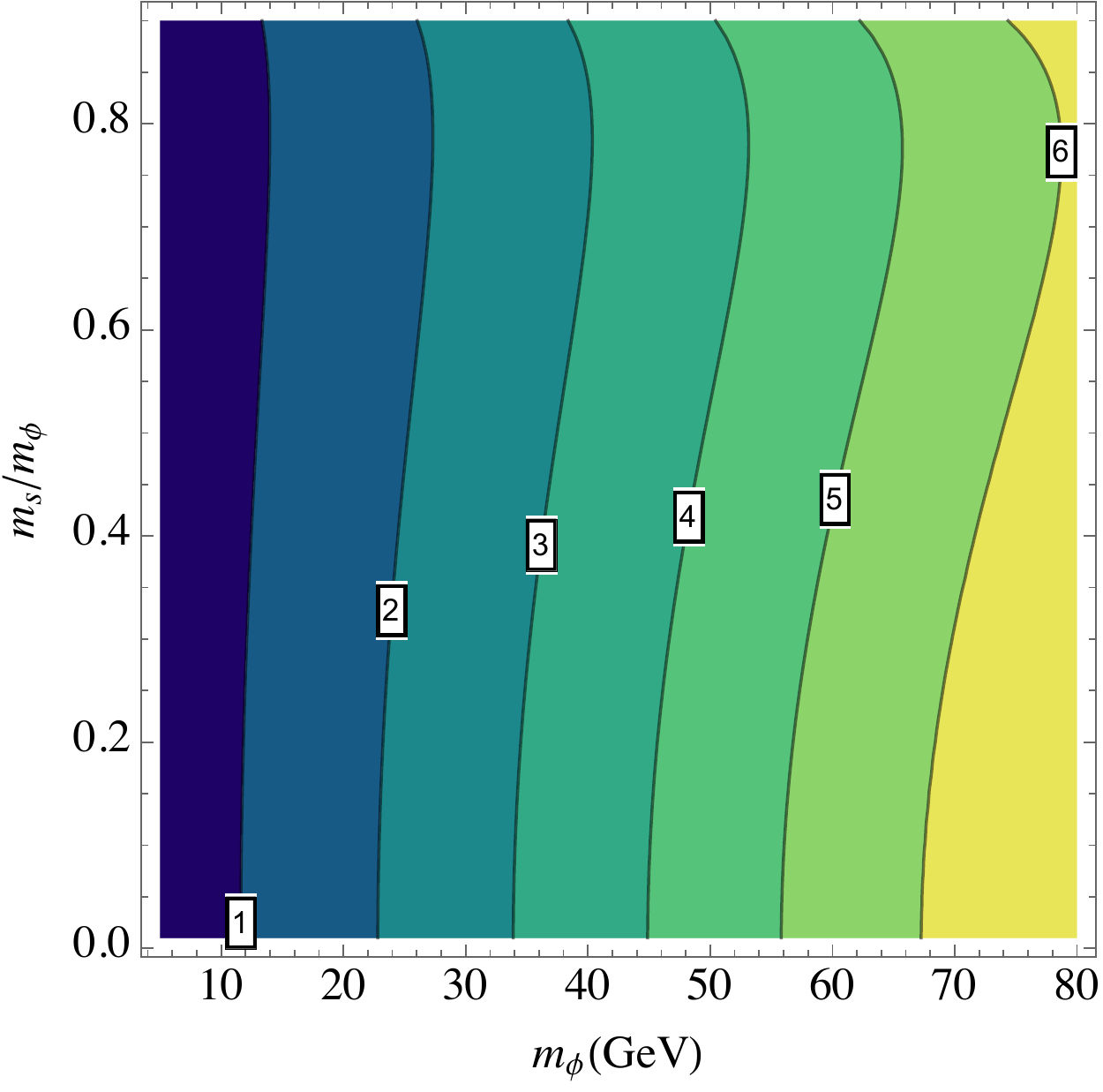}
\hspace{20mm}
\includegraphics[height=65mm]{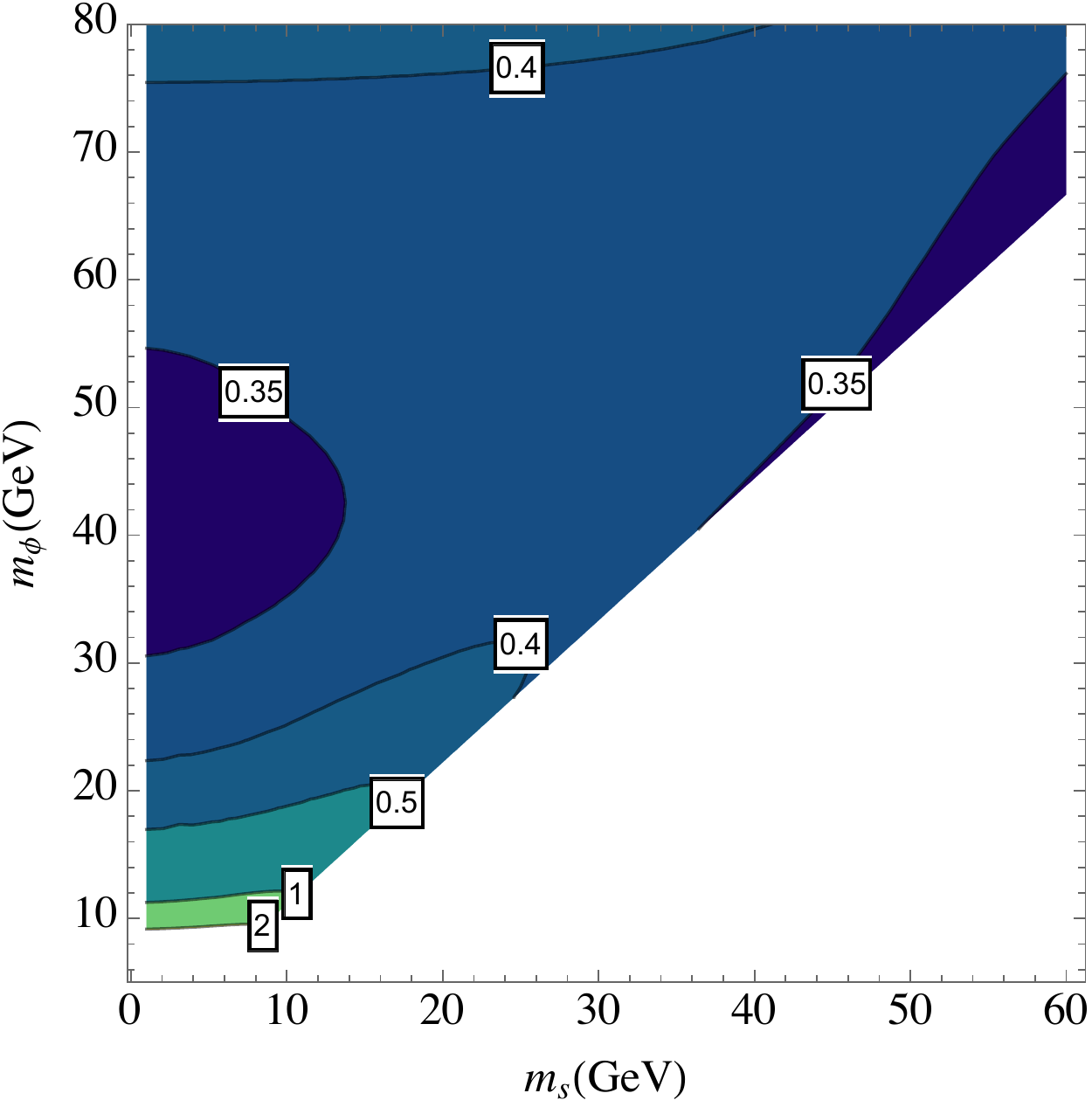}
\caption{Left: The quartic coupling
  $\lambda_4$ needed to obtain $\Omega h^2 = 0.11$ for complex scalar
  DM $\phi$ annihilating to pairs of dark scalars $s$ in the
  reference model of Eq.~(\ref{eq:scalarpot}), as a function of
  $m_{\phi}$ and $m_s = x m_\phi$ (using the approximated $s$-wave
  solution to the Boltzmann equation). Contours show the required
  value of $\lambda_4$ in units of $10^{-2}$.  Right: The maximum
  value of the Higgs mixing parameter $\epsilon$ allowed by LUX
  \cite{Akerib:2013tjd}, when the dark quartic coupling $\lambda_4$
  has been set to its thermal relic value.  In both panels we have set
  $y=1$.}
 \label{fig:scalarthermal}
\end{center}
 \end{figure*}
%%%%%%%%%%%%%%%%%%%%%%%%%%%%%%%%%%%%

Fixing the dark coupling constant to yield the thermal relic
abundance, direct detection experiments then constrain the coupling
of the dark vector to the SM.  The direct detection cross-section per
nucleon is
\begin{equation}
\label{eq:DDV}
\sigma_{\chi n}  =\frac{g_D^2 \bar g_N^2}{\pi m_{\chi}^4} \mu^2,
\end{equation}
where $\mu$ is the DM-nucleon reduced mass, and $\bar
g_N$ is the average nuclear coupling,
\beq
\bar g_N^2 \equiv \left(\frac{Z g_p + (A-Z)g_n}{A}\right)^2.
\eeq
In Eq.~(\ref{eq:DDV}), we have kept the leading
  contribution from $V_D$ exchange only, neglecting the subleading
  contributions from $Z$ exchange, which are suppressed by
  $\mathcal{O}(m_V^2/m_Z^2)$.  This is a good approximation in the
  best-fit regions for the vector model.

In the right panel of Fig.~\ref{fig:alphaDthermal}, we plot the
maximum allowed $\epsilon$ as a function of DM and vector
masses.  To generate this plot we have used the approximate analytic
$s$-wave relic abundance calculation to determine the thermal value of
$\alpha_D$. 
For a given coupling $\alpha_D$ appropriate to obtain the observed DM
relic density, as determined in Fig.~\ref{fig:alphaDthermal} (left),
one can satisfy the LUX bound by taking $\epsilon$ smaller than the
(upper bound) contours of Fig.~\ref{fig:alphaDthermal} (right). The
values of $\epsilon$ required to satisfy LUX limits still allow the
dark vector to decay long before BBN.

The cosmological history of the dark
sector also depends on $\epsilon$ weakly through the temperature
$T_{\rm dec}$ when $\chi$ and the SM depart from kinetic equilibrium.  The
kinetic mixing sets the strength of interactions of the form $\chi
f\to \chi f$, and for $\epsilon \ll 1$, DM-SM scattering may drop
below the Hubble expansion.  Once this kinetic decoupling occurs, SM
species leaving the thermal plasma deposit their entropy only into
lighter SM particles, and the SM is accordingly hotter than the dark
matter.  Setting $n_f \sigma v (\chi f \to \chi f) \lesssim H(T)$, we
estimate
\begin{align}
T_{\rm dec}^2 \approx &~(10 \mathrm{\,MeV})^2 \left(\frac{0.003}{\alpha_D}\right)\left(\frac{m_V}{10\,\mathrm{GeV}}\right)^4\\
&\notag
\hspace{10mm}\times
   \left(\frac{20 \,\mathrm{GeV}}{m_\chi}\right)\left(\frac{10^{-4}}{\epsilon}\right)^2 \left(\frac{4}{\sum g_i Q_i^2}\right)~,
\end{align}
where the sum runs over the relativistic visible sector states carrying charge $Q$, with internal degrees of freedom $g_i$, but for the case at hand this is simply the electron.
Thus $\chi$ stays in kinetic equilibrium with the SM well below the
freeze-out temperature in our best-fit region, provided $\epsilon$ is
not too far below current bounds.  Accordingly, in our calculation
above we have for definiteness taken $T_{\rm DM} = T_{\rm SM} = T$.  For
$\epsilon \lesssim 10^{-6}$, a complete treatment of kinetic
decoupling would be required for a precise prediction of the
present-day annihilation cross-section.  So long as kinetic decoupling
occurs prior to the chiral phase transition, these corrections are
small in comparison to the current astrophysical uncertainties.

There are many experimental constraints on kinetically mixed $U(1)$s;
see \cite{Essig:2013lka} for a summary.  For dark vector masses $m_V
>10$ GeV, the most stringent constraints come from electroweak
precision measurements, and limit $\epsilon \lesssim 0.02$ for most
values of $m_V$ below the $Z$ pole \cite{Hook:2010tw}.
For dark vectors lighter than $m_{\Upsilon}\approx 9.5$ GeV there are
more stringent limits from invisible quarkonium decay
\cite{Aubert:2009ae}, constraining the mediator couplings to be
$\epsilon \lesssim 10^{-3}$.  These bounds do not significantly
constrain the region of parameter space that is allowed by LUX.

This leaves a broad swath of $m_V$ and $\epsilon$ that are allowed by
all collider experiments and cosmological bounds, and are capable of
explaining the GC excess.
The best prospects for discovering this model at the LHC
would be through the exotic decay of the SM Higgs to the new vector
boson, either in association with a $Z$ boson, $h\to ZZ_D$, or in
pairs, $h\to Z_D Z_D$, depending on the structure of the dark
symmetry-breaking sector \cite{Curtin:2013fra}.

%%%%%%%%%%%%%%%%%%%%%%%%%%%%%%%%%%%%%%%%%
\section{Cascade annihilation to scalar mediators}
\label{S4}
%%%%%%%%%%%%%%%%%%%%%%%%%%%%%%%%%%%%%%%%%

In this section we consider simple reference models for DM
annihilating to pairs of states which couple to the SM through the
Higgs portal, i.e., have couplings to SM states proportional to their
Yukawa couplings (at leading order). The simplest such model that
allows for a present-day thermal cross-section of the correct order of
magnitude is scalar DM annihilating to pairs of scalars.  Fermionic DM
annihilating to pairs of scalars, through DM-scalar annihilations
alone, does not have any $s$-wave thermal cross-section, and is
therefore not a compelling explanation for the observed GC excess.

We consider a simple Higgs portal scenario
\cite{Patt:2006fw,MarchRussell:2008yu}.  We take the DM $\phi$ to be a
complex scalar and let it carry a global $U(1)$ quantum number
guaranteeing its stability (taking $\phi$ real and stabilized by a
$Z_2$ parity would not materially affect our conclusions).  We
consider the interactions between $\phi$, $s$, and $h$ to be given by
\beq
V(\phi, s, H) = &V(|\phi|^2) + V(|H|^2)+ \frac{\lambda_4}{2} |\phi | ^ 2 s^2 \\
  &+ \frac{\epsilon}{2} s^2 |H | ^ 2 
 + \left(-\frac{\mu_s^2}{2} s^2 + \frac{\lambda_s}{4!}s^4
\right) .
\label{eq:scalarpot}
\eeq
Suppose that the VEV $\langle s \rangle$ spontaneously breaks
the $Z_2$ parity $s\to -s$ of the potential.  After EWSB and $Z_2$
breaking, the mixing angle between $s$ and $h$ is then
\beq
\tan 2\theta \approx 2\theta =\frac{2\epsilon\langle s \rangle v}{m_h^2-m_s^2};
\eeq
we take $\theta \ll 1$ and hence the physical masses $m_{s,h}^2$ are
approximately equal to their unperturbed values.
In general the interactions of Eq.~(\ref{eq:scalarpot}) will induce
the interaction $|\phi|^2|H|^2$ at loop level; this will be negligible
in comparison to the tree-level interactions in the regime of
interest.

The independent parameters of this simple scalar reference model may
be taken as $m_\phi$, $\epsilon$, $\langle s \rangle$, $m_s$, and $\lambda_4$.
Thus, compared to the dark vector model, we have one additional
parameter: $y \equiv \langle s \rangle/m_s$, which we generically expect to be
$\mathcal{O}(1)$.  The thermal cross-section can be written 
\begin{eqnarray}
\langle\sigma v\rangle &=& \frac{1}{64\pi} \frac{\sqrt{1-m^2_s/m^2_\phi}}{m_\phi^2}
  \lambda_4^2 
  \times \\
\nonumber
  &&  \left(1+\frac{3 m_s^2}{4 m_\phi^ 2-m_s^ 2}-\frac{2\lambda_4\langle s \rangle^2}{2 m_\phi ^ 2-m_s^2}\right)^2 + \mathcal{O}(v^2).
\end{eqnarray}
Note that the dependence on $y$ enters only in the last term in
parentheses, multiplying $\lambda_4$.  Since we will find that
$\lambda_4\sim 10^{-2}$ to achieve the needed relic abundance, this
renders the thermal relic abundance insensitive to $y$.  In
Fig.~\ref{fig:scalarthermal} (left) we show the value of $\lambda_4$
needed to yield the correct thermal relic abundance, as obtained using
the approximate analytic $s$-wave solution to the Boltzmann equation.
Exact solution of the Boltzmann equation predicts present-day thermal
cross-sections again in the range 
\beq
\left. \langle \sigma v\rangle\right|_{\mathrm{today}} = (4.3-4.5)\times 10^{-26}\mathrm{cm}^3/\mathrm{s}.
\eeq

The leading contribution to direct detection is single $s$ exchange,
giving the per-nucleon cross-section 
\beq
\sigma_{\phi N} = \frac{\mu^2}{4\pi m_\phi ^ 2} f_N^2 \lambda_4^2 \epsilon^2 \left(\frac{\langle s \rangle^4}{m_s^4}\right)   
 \left(\frac{m_N}{m ^ 2_h-m ^ 2_s}\right)^2~,
\eeq
where $m_N\approx0.94$ GeV is the nucleon mass.  
Again, we drop the subleading contribution from $h$ exchange, as it
  is higher order in $m_s^2/m_h^2$, and thus small in the region of
  most interest for the Fermi excess.  Note the DM-nucleon scattering is
proportional to $y^4$; thus the $y$ dependence allows additional
freedom to enhance or suppress the direct detection cross-section
relative to signals from the GC or from the LHC.  Fixing $y=1$, we
take the effective Higgs coupling to nucleons as $f_p = f_n = 0.345$
\cite{Cline:2013gha}\footnote{See Ref.~\cite{Crivellin:2013ipa} for an
  alternate approach to Higgs-nucleon couplings.}, and show the
resulting constraints on the Higgs mixing $\epsilon$ in
Fig.~\ref{fig:scalarthermal} (right).  Constraints from direct
detection are relatively weak, requiring $\epsilon \lesssim 0.4$ in
the best-fit regions of parameter space.

\begin{figure}[t!]
\includegraphics[height=70mm]{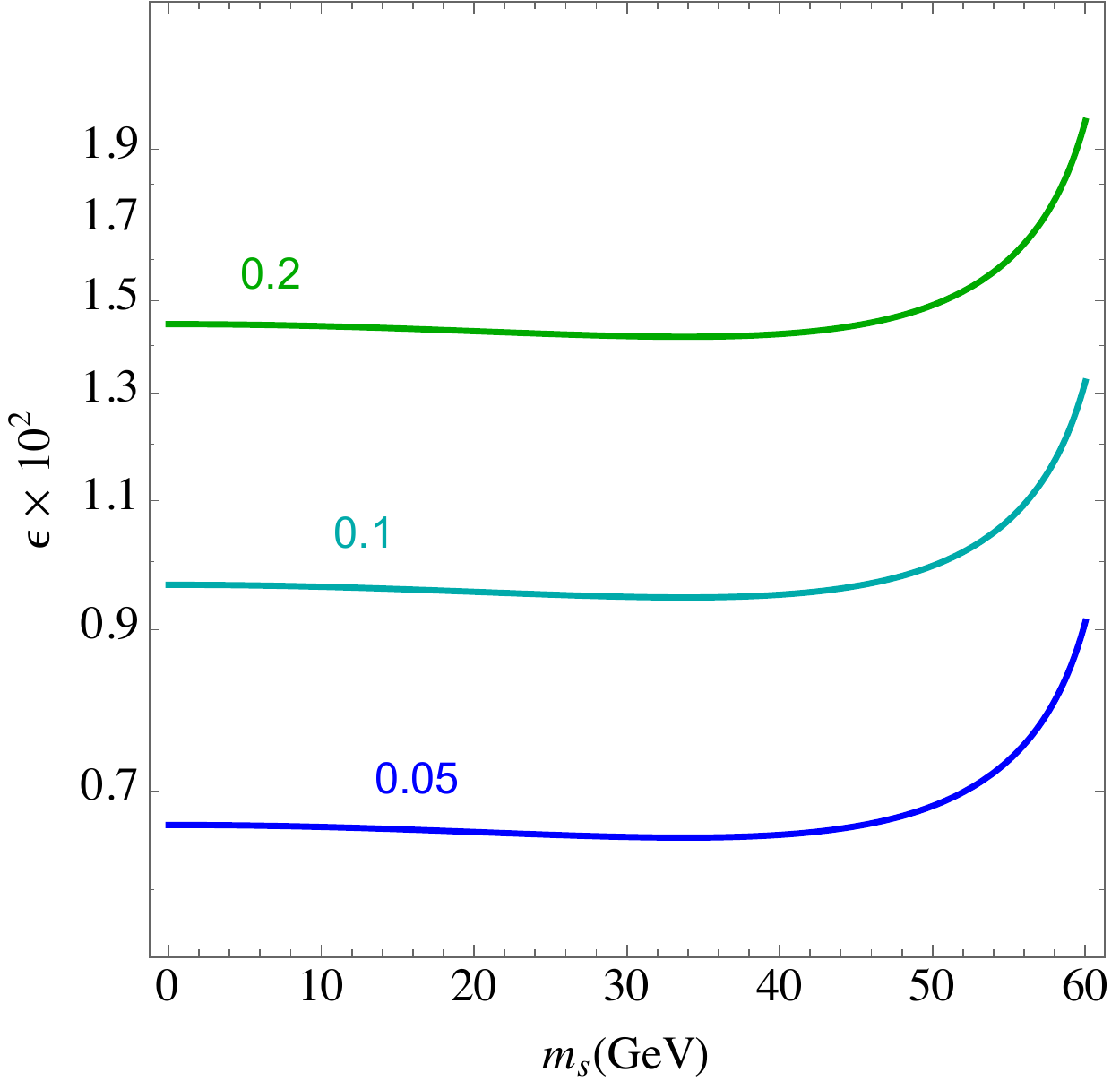}
\caption{Limits on the scalar-Higgs mixing parameter $\epsilon$ from
Higgs properties at the LHC, as a function of mediator mass $m_s$.   The three contours denote $\mathrm{Br} (h\to s s ) = 0.2$,
$0.1$, and $0.05$, as indicated.
\label{fig:exoHconstraint}}
\end{figure}

 More stringent constraints come from LHC
limits on the total non-SM branching fraction of the Higgs, which
limit $\mathrm{Br} (h\to \mathrm{BSM})\lesssim 0.2$ if SM production
is assumed
\cite{Belanger:2013kya,Giardino:2013bma,Ellis:2013lra}.
In the mass ranges of interest for the GC excess, the scalar mass satisfies $m_s<m_h/2$.
Thus the scalar potential of Eq.~(\ref{eq:scalarpot}) yields
a partial width for $h\to s s$,
\beq
\Gamma (h\to s s) = \frac{\epsilon^2 v^2}{32\pi m_h}\,
\left(\frac{m_h^2+2m_s^2}{m_h^2 - m_s^2}\right)\sqrt{1-\frac{4 m_s^2}{m_h}}.
\eeq
Requiring
$\mathrm{Br} (h\to s s) < 0.2$ restricts $\epsilon \lesssim 1.5\times 10^{-2}$;
future measurements of the Higgs sector with $5$-$10\%$ precision~\cite{Peskin:2013xra} can be sensitive to $\epsilon \gtrsim
0.7\times 10^{-3}$, as we show in Fig.~\ref{fig:exoHconstraint}.  Values of
$\epsilon \sim 10^{-3}$ correspond to $\theta \sim 10^{-4}$. Such
values of $\theta$ still allow $s$ to decay well before BBN in the
mass ranges under consideration.  As in the dark vector case, for
$m_{s}\lesssim 10$ GeV there are additional limits from invisible quarkonium
decay, but these are typically less constraining than the
inclusive Higgs decay limit, giving limits of $\epsilon\lesssim
10^{-2}$ \cite{Essig:2013vha}. 

The small Yukawa couplings of the light SM fermions imply in general
that for $\epsilon \lesssim 10^{-3}$, kinetic decoupling occurs at
temperatures of the same order as the freeze-out temperature.  Thus a
full numerical solution of the Boltzmann equation would be required to
precisely determine the relic abundance for a given $\lambda_4$ and
$\epsilon$.  As kinetic decoupling again occurs prior to the chiral
phase transition, the net impact on the deduced dark matter properties
is small relative to the astrophysical uncertainties, and the
approximate treatment presented here is sufficient to assess the
relations between indirect, direct, and collider signals expected in
this model.

%%%%%%%%%%%%%%%%%%%%%%%%%%%%%%%%%%%%
%
%%%%%%%%%%%%%%%%%%%%%%%%%%%%%%%%%%%%

For $\epsilon$ values consistent with $\mathrm{Br} (h\to
\mathrm{BSM})\lesssim 0.2$, $s$ will be difficult to see at
colliders. Direct observation of the exotic Higgs decays $h\to s s$,
with $s$ decaying according to the SM Yukawa couplings, will be
challenging at the LHC; see \cite{Curtin:2013fra} for a recent survey
and review.  The best prospects exist in the region $m_s < 2 m_b$,
where the enhanced ${\rm Br}(s\to\tau\tau)$ may allow a detection for
${\rm Br}(h\to s s)\gtrsim 5\%$; for $m_s > 2 m_b$, sensitivity to
$\mathrm{Br}(h\to s s )\gtrsim 20\%$ is a reasonable target
\cite{Curtin:2013fra}.

%%%%%%%%%%%%%%%%%%%%%%%%%%%%%%%%%%%%%%%%%%%%%%%%%%%%%%%%
\section{Cascade annihilations to Gluophilic mediators}
\label{S5}
%%%%%%%%%%%%%%%%%%%%%%%%%%%%%%%%%%%%%%%%%%%%%%%%%%%%%%%%

Let us re-examine the scalar DM model of Sect.~\ref{S4} in the case
that $\langle s\rangle=0$. To allow $s$ to decay, we add to
the spectrum a pair of heavy vector-like fermions $\psi$ in the SM
representation $(3,1)_Y$ + h.c.~which are coupled to the DM as follows:
\beq
\mathcal{L}\supset & \frac{\lambda_4}{2} |\phi | ^ 2 s^2 
+ \lambda_{\psi} s \psi\bar \psi - M \psi\bar \psi~.
\eeq
Integrating out these heavy exotic fermions leads to 
\beq
\mathcal{L}_{\rm eff}\supset & \frac{\lambda_4}{2} |\phi | ^ 2 s^2 
+ \frac{1}{\Lambda}s\, G^a_{\mu\nu}G^{a\mu\nu}~.
\eeq
The mass scale $\Lambda$ can be identified with~\cite{Chang:2005ht}:
\beq
\Lambda^{-1} = \frac{\alpha_S\,b_i\,\lambda_{\psi}}{8\pi M}.
\eeq
where $b_i$ is the $\beta$-function contribution due to $\psi$, 
e.g.~$2/3$ for an SU(3) triplet. This scenario allows
for cascade annihilations of the form $\phi \phi^\dag \rightarrow s
s$, and subsequently $s \rightarrow gg$. The resulting gamma-ray
spectrum for this case is studied in Fig.~\ref{figE} (right).  The
decay rate of the scalar state to gluons is
\beq
\Gamma_s =  \frac{8\,m^3_s}{4\,\pi\,\Lambda^{2}}~.
\eeq
Decays before BBN occur provided the coloured fermions
have mass $M/\lambda_\psi \lesssim 10^{11}$ GeV.
This coupling via exotic heavy fermions is analogous to KSVZ-type
axion models \cite{Kim:1979if,Shifman:1979if}.

For $Y\neq0$ the heavy fermion loop also induces the operator $s\,
F_{\mu\nu}F^{\mu\nu}$ and thus also the decay
$s\rightarrow\gamma\gamma$.
This decay directly to photons would result in a gamma-ray line for
$s$ at rest in the galactic frame.  In the case of interest for us,
where $x=m_s/m_\phi \sim 0.5$, the boost of the $s$ particles relative
to the galactic frame widens the line into a box \cite{Ibarra:2012dw}:
\begin{eqnarray}
\nonumber
\frac{1}{N_\gamma} \frac{dN_\gamma}{dE_\gamma} &=& \frac{1}{m_\phi \sqrt{1-x^2}}
     \Theta \left(E_\gamma - \frac{m_\phi}{2}(1-\sqrt{1-x^2})\right) \times\\
&& 
 \Theta \left(\frac{m_\phi}{2}(1+\sqrt{1-x^2}-E_\gamma)\right).
\end{eqnarray}
The intensity of the gamma-ray box relative to the gluon-induced
continuum contribution will be suppressed by the factor
$\left(\alpha_{\rm em}/\alpha_s\right)^2$, and is not currently
observable for the parameters of interest \cite{Ibarra:2012dw}.  On
the other hand, the box extends to significantly higher photon
energies $E_\gamma$ and has a kinematic feature at the upper
endpoint, and thus presents an interesting target for future
experiments.

\begin{figure}[t!]
\includegraphics[height=70mm]{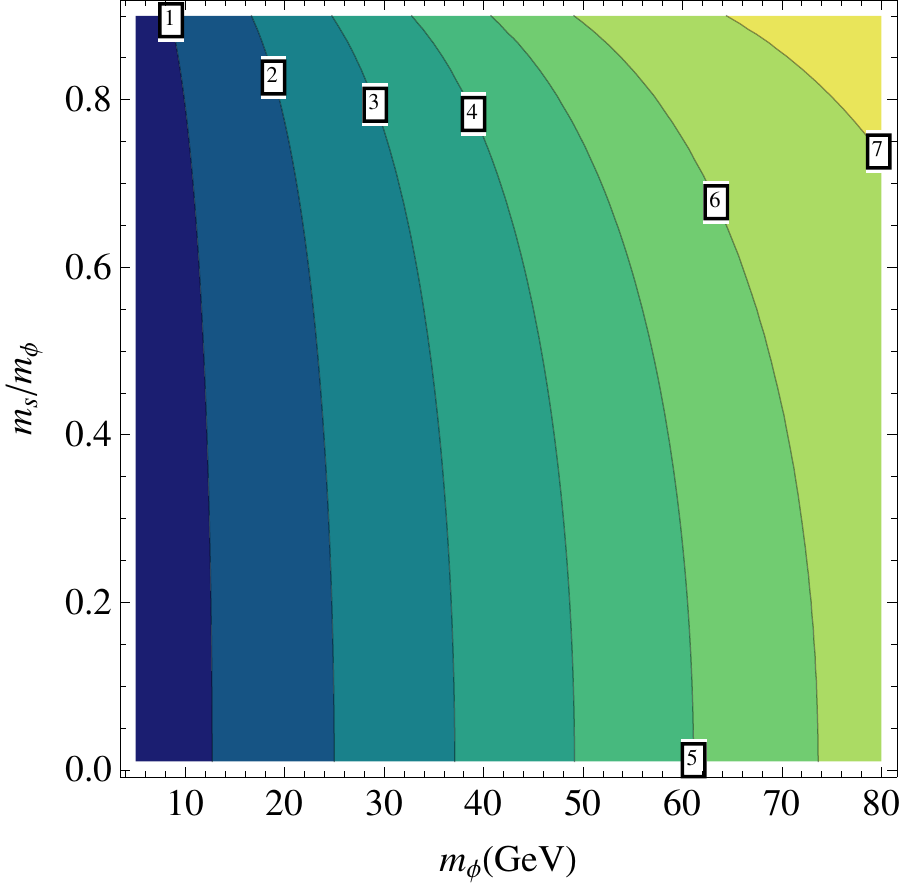}
\caption{The quartic coupling
  $\lambda_4$ needed to obtain $\Omega h^2 = 0.11$ for complex scalar
  DM $\phi$ annihilating to pairs of dark scalars $s$ in the
   model of Sect.~(\ref{S5}), as a function of
  $m_{\phi}$ and $m_s = x m_\phi$ (using the approximated $s$-wave
  solution to the Boltzmann equation). Contours show the required
  $\lambda_4$ in units of $10^{-2}$. 
 \label{fig:phiphitoss}}
  \end{figure}

Similar to the previous cases, in Fig.~\ref{fig:phiphitoss} we show
the value of $\lambda_4$ required to obtain the observed relic
density. These results apply also to DM annihilations to pairs of
pseudo-scalars, $\phi\phi\to a a$. For
definiteness, we take the dark sector and the SM to be reheated 
to the same temperature $T_{\rm reheat}>m_{top}$, but kinetically decoupled
thereafter.  
As DM freeze-out occurs prior to the chiral phase transition in 
our best-fit regions ($m_{\rm DM}\sim40$ GeV), 
we approximate $T_{\rm DM} = 0.83 T_{\rm SM}$.

%%%%%%%%%%%%%%%%%%%%%%%%%%%%%%%%%%%%
%
%%%%%%%%%%%%%%%%%%%%%%%%%%%%%%%%%%%%

The leading direct detection cross-section is spin-independent and
highly suppressed. Parameterizing the effective $s$-nucleon coupling
as \cite{Cheng:2012qr}\footnote{The accuracy of the
    SU(3) chiral perturbation theory used to extract the gluonic
    matrix elements $\langle N|GG|N\rangle$ and $\langle
    N|G\tilde{G}|N\rangle$ has been questioned, and an alternate approach has
    recently been proposed for the scalar matrix element $\langle |m_f
    f \bar f| \rangle$ in Ref.~\cite{Crivellin:2013ipa}; however, as our
    aim here is to establish a parametric estimate, the precision is
    more than sufficient for our purposes.}
\beq
g_{sNN} \simeq \frac{\langle N|G G|N\rangle}{\Lambda} \simeq -\frac{4\pi}{\alpha_s\Lambda}
  \times 180\mathrm{\,MeV}~,
\eeq
direct detection proceeds at one-loop, with cross-section parametrically
given by
\beq
\sigma_{\phi N}\sim \frac{1}{16\pi} \frac{1}{(m_\phi+m_N)^2}
      \frac{\lambda^2_4 g_{sNN}^4}{16\pi^2} \left(\frac{m_N}{m_s}\right)^4.
\eeq
Clearly, for $M/\lambda_\psi\gg$ TeV, there is no hope of a direct
detection signal.

Similar results hold for pseudo-scalars $a$, although $\Lambda$ is a factor of 3 larger.
In this case, we
estimate the effective $a$-nucleon coupling as \cite{Cheng:2012qr}
\beq
g_{aNN} \simeq \frac{\langle N|G \widetilde G|N\rangle}{\Lambda} \simeq -\frac{8\pi}{\alpha_s\Lambda}
  \times \left\{\begin{array}{cl} -380 \mathrm{\,MeV} & p \\ 10 \mathrm{\,MeV} & n \end{array} \right.
\eeq
The leading direct detection cross-section is spin-independent and
highly suppressed, with similar parametric dependence to the scalar
case.

%%%%%%%%%%%%%%%%%%%%%%%%%%%%%%%%%%%%%%%%%%%%%%%%%%%%%%
\section{Concluding remarks}
\label{S6}
%%%%%%%%%%%%%%%%%%%%%%%%%%%%%%%%%%%%%%%%%%%%%%%%%%%%%%

If its origin from DM is confirmed, the GC gamma-ray excess will not
only present the long-awaited detection of DM, but also provide
evidence of a dark sector governing its interactions.  The DM mass and
thermal cross-section indicated by the excess is highly suggestive of
a particle of mass $m_{\rm DM}\sim $ tens of GeV, freezing out via
thermal annihilations mediated by a BSM particle, with a substantial
component of the annihilation proceeding through the $s$-wave.  For DM
with this mass and this magnitude of coupling, the big challenge for
models of DM is to reconcile the signal from the GC with the lack of
any signal at direct detection experiments, especially LUX.

The deduced need for a BSM mediator naturally presents an elegant
solution to this puzzle.  When $m_{\rm med} < m_{\rm DM}$, the
mediator itself is available as a final state for DM annihilations.  A
small coupling of the mediator to the SM subsequently allows the
mediator to decay to SM particles with a cosmologically short
lifetime.  Thus in this scenario the coupling of the dark sector to SM
particles in general and to nucleons in particular is no longer tied
to the thermal relic density of DM.  This cascade annihilation
scenario opens up the scope of possible masses and quantum numbers for
both DM and the mediator.

The photon spectra arising from cascade annihilations will be
determined by the masses of the DM and mediator, together
with the details of how the mediator decays.  We have studied three
well-motivated minimal scenarios for mediator decay: vector mediators
decaying to the SM via hypercharge mixing; scalar mediators decaying
to the SM through Higgs mixing; and (pseudo)-scalar mediators decaying
to the SM via loops of heavy fermions.  For all scenarios we establish
the combinations of masses that lead to the best description of the
gamma-ray spectrum, as determined in \cite{Daylan:2014rsa}.

We construct simple reference models for all of these scenarios and
demonstrate that they are capable of explaining the GC excess whilst
naturally eluding direct detection as well as collider limits.  Our
best-fit scenarios are compared to the observations in
Fig.~\ref{figF}. There is a large and natural range of parameter space
in all of these models capable of explaining the excess.  This ease in
evading current experiments has the unfortunate consequence that
signals of these models at direct detection and at colliders are
naturally small, and may be difficult to observe at the LHC. Exotic
decays of quarkonia and of the SM Higgs boson offer the best prospects
for terrestrial experiments.  Astrophysical signals offer a window
onto the complementary regime where the mediator has small couplings
to the SM, via perturbations of BBN or the CMB via late-decaying
particles.

We have focused our attention on minimal models, with one DM particle
and one mediator species active in the process of thermal freeze-out.
Less minimal choices of mediator-SM couplings may also be constructed.
For instance, hidden vectors may be constructed that couple to the SM
through leptophobic \cite{Dobrescu:2014fca,Babu:1996vt,Tulin:2014tya}
or leptophilic \cite{Fox:2008kb} portals.  In such cases the DM
thermal freeze-out will generically proceed as computed in
Sect.~\ref{S4}, but the constraints on the mediator couplings to the
SM, as well as limits on the additional matter species that appear in
such models, will impose different restrictions on the allowed
parameter space.
Further, a much broader range of possible DM quantum numbers and
interactions opens up if the mediator sector is allowed to be
non-minimal, enabling such possibilities as mixed annihilations $XX\to
\phi_1\phi_2$ \cite{Nomura:2008ru}, longer cascades
\cite{Mardon:2009rc}, or showering in a hidden sector.  There are many
opportunities for extending DM model building in these directions.

%%%%%%%%%%%%%%%%%%%%%%%%%%%%%%%%%%%%%%%%%%%%%%%%%%%%%%%%%%%%%%%

\section*{Acknowledgements}

We are grateful to Dan Hooper, Robert Lasenby, Tim Linden, Torbj\"orn
Sj\"ostrand, and Tracy Slatyer for useful discussions. Note: Preprints
\cite{Abdullah:2014lla,Boehm:2014bia,Ko:2014gha}, which appeared in
the process of finalising this work, consider complementary aspects
 of DM cascade annihilations.

%%%%%%%%%%%%%%%%%%%%%%%%%%%%%%%%%%%%%%%%%%%%%%%%%%%%%%%%%%%%%%%

\end{document}